\documentclass[mathleft
]{an}
\usepackage{graphicx}
\usepackage{times}
\begin{document}

% The following seven commands are intended for editorial usage and should be ignored by
% the author(s).
\Pagespan{789}{}% Document's page range.
% If second parameter is left empty, the last page is computed automatically.
\Yearpublication{2010}%
\Yearsubmission{2010}%
\Month{04}%
\Volume{999}%
\Issue{88}%
% \DOI{This.is/not.aDOI}%

\title{HD~207331 a new $\delta$ Scuti star in the Cygnus field: discovery and follow-up observations  -- \\
        }

\author{L. Fox Machado\inst{1}\fnmsep\thanks{Corresponding author:
  \email{lfox@astrosen.unam.mx}\newline}
%Example
%for footnote, note the usage of the \texttt{fnmsep}
%command as separator between institute number and footnote mark}
, W.~J. Schuster\inst{1}, C. Zurita\inst{2}, J.~S. Silva\inst{1}, R.
Michel\inst{1}}
\titlerunning{HD 207331 new $\delta$ Scuti variable}
\authorrunning{L. Fox Machado et al.}
\institute{Observatorio Astron\'omico Nacional, Instituto de
Astronom\'{\i}a -- Universidad Nacional Aut\'onoma de M\'exico, Ap.
P. 877, Ensenada, BC 22860, M\'exico \and Instituto de
Astrof\'{\i}sica de Canarias, C\ V\'{\i}a L\'actea s/n, 38205, La
Laguna, Tenerife, Spain}

\received{March 31, 2010}
\accepted{April 1, 2010}
\publonline{later}


\abstract{%
  Preliminary results on the discovery and follow-up observations
  of a new $\delta$ Scuti pulsator in the
  Cygnus field are presented. The variability of the star HD 207331 was
  detected while testing a Str\"omgren spectrophotometer attached to
  the H.~L. Johnson 1.5-m telescope at the San Pedro M\'artir observatory,
  M\'exico.  CCD photometric data acquired soon after confirmed its variability.
  A few hours of $uvby$ differential photoelectric photometry during
  three nights revealed at least two beating periods.
  A two-site observational campaign carried out during one week in
  2009 confirms the multi-periodic nature of this new $\delta$ Scuti
  pulsator.}

\maketitle

\section{Introduction}

The $\delta$ Scuti-type pulsators  are stars with masses between 1.5
and 2.5 $M_{\odot}$ located at the intersection of the classical
Cepheid instability strip with the main sequence. They have spectral
types A and F, a period range between 0.5 h to 6 h, and generally
pulsate with a large number of radial and nonradial modes excited by
the $\kappa$ mechanism. This  makes them interesting targets for
seismic studies. Therefore, any new detection of a $\delta$ Scuti
star can be a valuable contribution to asteroseismology.

Since most of the $\delta$ Scuti stars are short period variables
with typical photometric amplitude of 20 mmag, their oscillations
can be  easily detected from the ground.  In fact, several $\delta$
Scuti stars have been discovered accidentally when taken as
reference stars of observations of well known $\delta$ Scuti stars
(e.g. Fox Machado et al. 2002 and 2007; Li et al. 2002). Others have
been catalogued either in surveys devoted to the characterization of
new variables or, as in this case, when considered as
constant stars while testing observatory equipment.  In particular,
this paper presents a summary of the observations which yielded  the
discovery and characterization of the new $\delta$ Scuti star, HD
207331.

\begin{figure*}
\includegraphics[width=15cm,height=17cm]{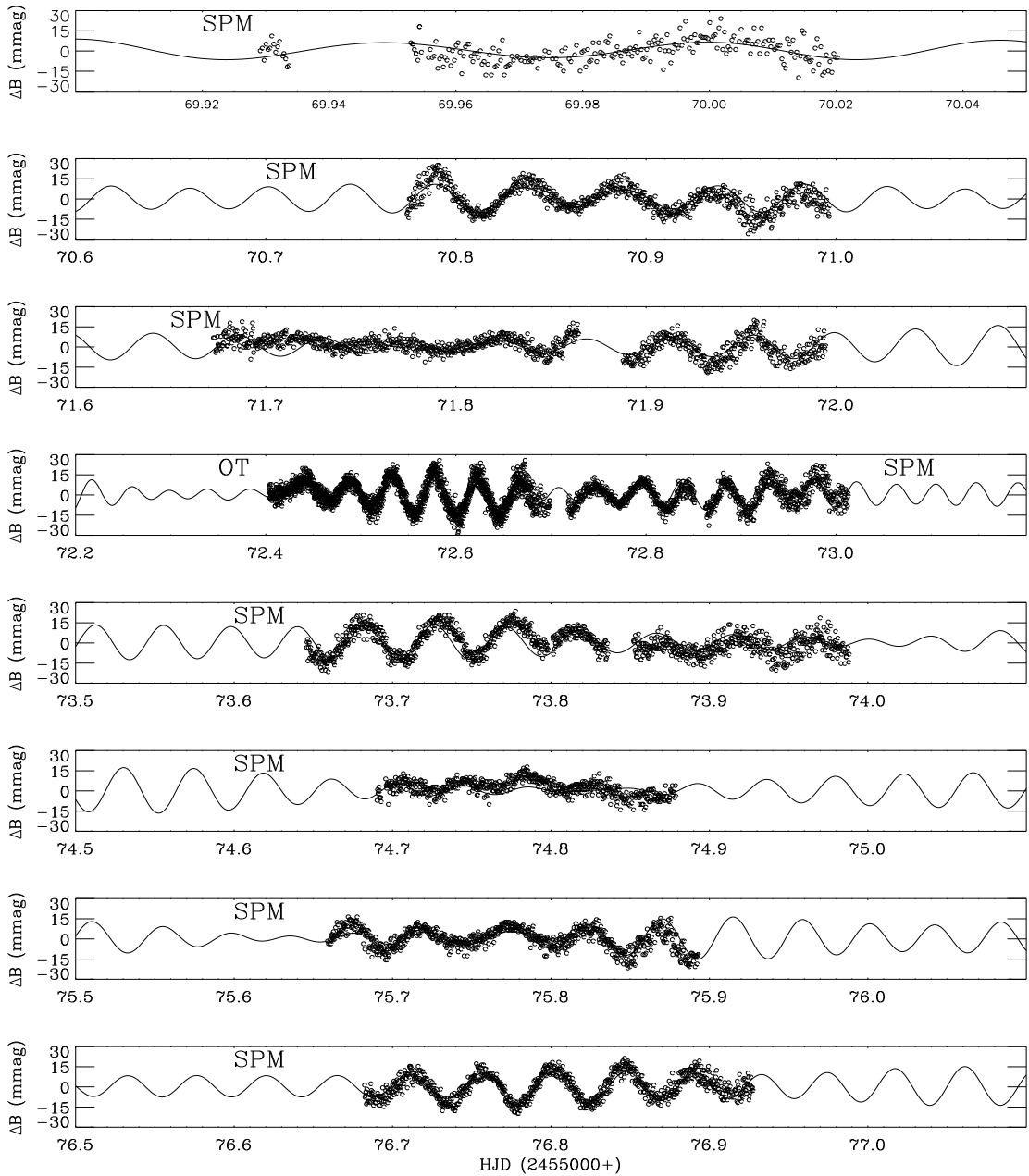}
\caption{CCD differential light curve:  HD 207331 - Comparison. The
fit of the four-frequency solution to the data is shown by solid
line.} \label{fig:curves}
\end{figure*}

\section{Observations}
\subsection{Discovery}
The star  HD~207331 ($=$ SAO 51294, BD$+$42 4207, HIP 107557) was
observed in a sample of A-type stars during a few hours on the night
of September 27, 2007, in order to test the six-channel $uvby$--$\beta$
spectrophotometer attached to the H.~L. Johnson 1.5-m telescope of
the San Pedro M\'artir Observatory, Baja California, M\'exico. The
variability of the star was clear, despite the fact that it had been
classified in the SIMBAD database as a normal A0 star. CCD photometric
observations of HD 207331, confirming its variability, were carried
out on the night of September 30, 2007, with the 0.84-m telescope
(Schuster et al.~2008; Fox Machado et al.~2008).

\subsection{Follow-up observations}

\subsubsection{$uvby$ differential photometry}
$uvby$ differential photometry of HD 207331 was also performed
during three nights in November 2007 using the 1.5-m telescope and
the six-channel Str\"omgren spectrophotometer. In particular, the
star was monitored for about 3.5 h on November 11, for about 4 h on
November 18, and 3.5 h on November 19.  As a result of these
observations, a multiperiodic characteristic of the star, with at
least two beating periods, was found (for details see Fox Machado et
al. 2008).

Encouraged by results obtained in that short run and in order to
investigate its pulsation characteristics more accurately, we decided
to perform a two-site observational campaign on HD 207331 during
August/September 2009.

\subsubsection{Two-site observations}
The two observatories  involved in the observational campaign are listed
in Table~\ref{tab:tel} together with the telescopes and instruments
used. Table~\ref{tab:log} gives the log of observations.  Bad
weather conditions at the OT did not allow us to get more than one
night of data. However, a total amount of 53.8 hours of useful data
was obtained from the two sites.

\begin{table*}[!t]\centering
 \setlength{\tabcolsep}{1.0\tabcolsep}
 \caption{List of instruments and telescopes involved in the campaign.
Observers' abbreviations correspond to the initials of the co-authors.} \label{tab:tel}

  \begin{tabular}{lccc}
\hline
Observatory&  Telescope& Instrument & Observers  \\
\hline
 Observatorio del Teide (OT, Spain) &0.80m & 2048x2048 CCD & CZ \\
 Observatorio San Pedro M\'artir (SPM, M\'exico) &0.84m& 1024x1024 CCD&WJS, JS, LFM \\
\hline
\end{tabular}
\end{table*}

\begin{table*}
\caption{Log of observations.  Observing time is expressed in hours.}
\begin{tabular}{cccccr}

\hline

    Day& UT Date 2009 &  Start Time  & End Time &  OT  & SPM$\;$ \\
      &                &(HJD 2455000+)&(HJD 2455000+)&&       \\
 \hline
      1&    Aug 26    &    69.93     &   70.02  &  -   & 2.185 \\
      2&    Aug 27    &    70.77     &   71.00  &  -   & 5.335 \\
      2&    Aug 28    &    71.67     &   72.70  &7.086 & 7.731 \\
      4&    Aug 29    &    72.72     &   73.01  &  -   & 7.076 \\
      5&    Aug 30    &    73.65     &   73.99  &  -   & 8.218 \\
      6&    Aug 31    &    74.69     &   74.88  &  -   & 4.536 \\
      7&    Sep 01    &    75.66     &   75.89  &  -   & 4.592 \\
      8&    Sep 02    &    76.68     &   76.93  &  -   & 7.085 \\
    \hline
       &    Begin     &     End      &Total Time&  OT  & SPM$\;$ \\
       &    Aug 26    &    Sep 02    & 53.844   &7.086 &46.758 \\
\hline
\end{tabular}
\label{tab:log}
\end{table*}

\begin{table*}[!t]\centering
  \setlength{\tabcolsep}{1.0\tabcolsep}
 \caption{Position, magnitude, and spectral type of target, comparison,
and check stars observed in the CCD frame.}
  \begin{tabular}{lccccc}
\hline
Star       &       ID       &    RA     &   Dec      &   V   & SpTyp \\
           &                &  (2000.0) & (2000.0)   & (mag) &       \\
\hline
Target     &   HD 207331    & 21 47 02  &$+$43 19 19 &  8.3  & $A0$  \\
Comparison &  BD$+$42 4208  & 21 47 12  &$+$43 19 51 &  9.4  & $A0$  \\
Check      & TYC 3196-1243-1& 21 47 06  &$+$43 18 58 & 10.9  &  -    \\
\hline
\end{tabular}
\label{tab:stars}
\end{table*}

\begin{figure}
\includegraphics[width=8.0cm,height=6.0cm]{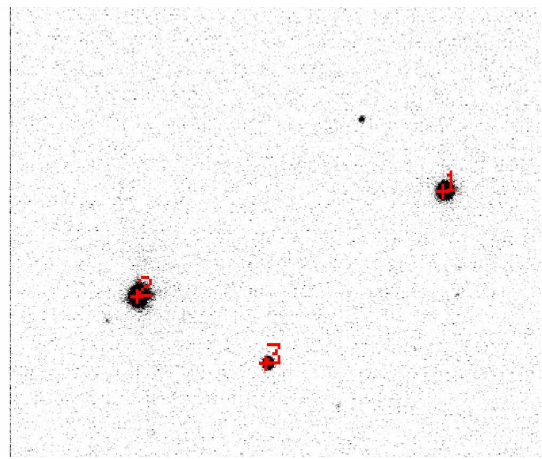}
\caption{The CCD FOV near HD 207331. 1 stands for the comparison
star,  2 for HD 207331, and 3 for the check star. Some properties of
the stars are listed in Table 1. North is up and East is right.}
\label{fig:field}
\end{figure}

\medskip
The observations were obtained through a Johnson $B$ filter.
Figure~\ref{fig:field} shows a section of a typical image of the
CCD's field of view ($10' \times 10'$) at the IAC-80 telescope of
the Teide observatory. The target star is labeled with number 2;
comparison and check stars with 1 and 3, respectively.
Table~\ref{tab:stars} shows the main observational parameters
corresponding to the target and comparison stars as taken from the
SIMBAD database operated by the CDS (Centre de Donn\'e\-es
astronomique de Strasbourg).

\medskip
 Sky flats,
dark and bias exposures were taken every \\
night at all sites. All data were calibrated and reduced using
standard IRAF routines. Aperture photometry was implemented to
extract the instrumental magnitudes of the stars. The differential
magnitudes were normalized by subtracting the mean of differential
magnitudes for each night. In Figure~\ref{fig:curves} the entire
light curves, HD 207331 - Comparison, are presented. As can be seen
from the fourth panel (from top to bottom) no overlapping of
the observations was obtained for the one night of observing at OT,
the 28th of August 2009.

\section{Period analysis}

The period analysis has been performed by means of standard Fourier
analysis and least-squares fitting. In particular, the amplitude
spectra of the differential time series were obtained by means of
Period04 package (Lenz \& Breger 2005), which utilizes Fourier as
well as multiple least-squa\-res algorithms. This computer package
allows us to fit all the frequencies simultaneously in the magnitude
domain.

\begin{figure}
\includegraphics[width=8.0cm,height=8.0cm]{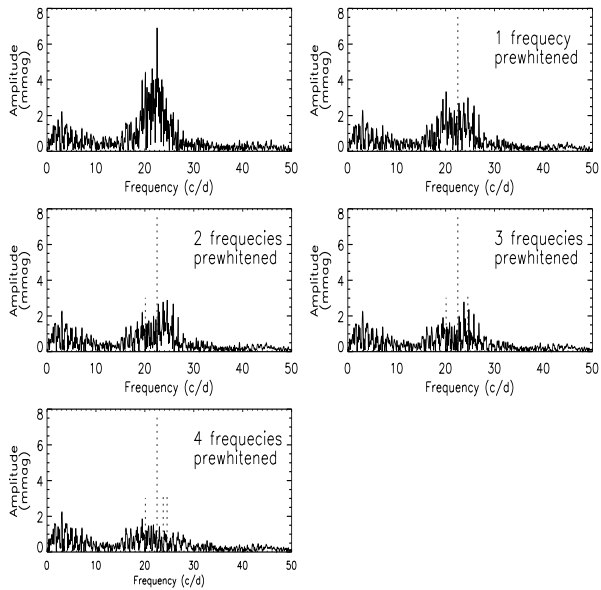}
\caption{Pre-whitening process in HD 207331. In each plot, from left
to right, the highest amplitude peak is selected and removed from
the time series, and a new spectrum is obtained.} \label{fig:spec}
\end{figure}

\medskip
The amplitude spectrum of the differential light curve, HD 207331 -
Comparison, is shown in the first plot of Fig.~\ref{fig:spec}. As
can be seen, HD 207331 shows high signal-to-noise\\ peaks around 22
cycle day$^{-1}$. The subsequent plots in the figure, from left to
right, illustrate the prewhitening process of the frequency peaks in
each amplitude spectrum.

\medskip
The frequencies have been extracted by means of a standard
prewhitening method. In order to decide which of the detected peaks
in the amplitude spectrum can be regarded as intrinsic to the star,
Breger's criterion has been followed (Breger et al.~2003),  where
it was shown that the signal-to-noise ratio (in amplitude) should be
at least 4 in order to ensure that the extracted frequency is
significant.

\medskip
The frequencies, amplitudes, and phases  are listed in
Table~\ref{tab:frec}. Four significant frequencies have been
detected in HD 207331. A comparison of this four-frequency solution
to the data is displayed in Fig.~\ref{fig:curves} with the solid line.

\medskip
The main oscillations peaks have been found in the 20--25 cycle
day$^{-1}$ range (i.e., $\sim$ 231.4--289.25 $\mu$Hz).  In
particular, the highest amplitude peak
is located at 22.49 cd$^{-1}$ (260.21 $\mu$Hz), and the next
significant frequency is located at 20.09 cd$^{-1}$ (232.47
$\mu$Hz ). As was found in the discovery data, two beating pulsation
modes are present in HD 207331, namely 24.54 cd$^{-1}$  and 23.74
cd$^{-1}$ . This is a common behavior in $\delta$ Scuti stars.

\begin{table}[!t]\centering
  \setlength{\tabcolsep}{1.0\tabcolsep}
 \caption{Frequency peaks detected in the light curve:  HD 207331 - Comparison.
S/N is the signal-to-noise ratio in amplitude after the prewhitening
process.} \label{tab:frec}
  \begin{tabular}{cccc}
\hline
    Freq.   &  A   & $\varphi$/($2\pi$)& $S/N$ \\
    (c/d)   &(mmag)&                   &       \\
\hline
   22.4880  & 7.76 &       0.18        &  13.7 \\
   20.0923  & 3.02 &       0.90        &   5.1 \\
   24.5384  & 3.22 &       0.32        &   6.3 \\
   23.7409  & 3.18 &       0.85        &   5.9 \\
 \hline
\end{tabular}
\end{table}

\section{Discussion and conclusion}
The determination of the evolutionary stage of a field star requires
precise estimates of its global parameters. In the case of HD 207331,
there is little information in the literature about its physical
parameters. In particular, the Hipparcos catalogue (Perryman et al.
1997) provides a parallax of 3.31 $\pm$ 0.88 mas, from which a
distance of 302 pc  can be estimated.  The large relative error
of the measured distance ($\sigma (\pi))/ \pi \sim 0.27$)  implies
$\sigma (M_V)\sim 0.6$ mag, making the Hipparcos absolute magnitude
very imprecise for HD 207331.

Therefore, with the information available it is not possible to make
a reliable seismic modeling for HD 207331. However, the complicated
oscillation spectrum of HD~207\-331, with two beating modes, points
to it being a fast rotating $\delta$ Scuti star. This would imply
that the four frequencies listed in Table~\ref{tab:frec} are due to
nonradial oscillations. In fact, as is well known, most of the
$\delta$ Scuti stars are rapid rotators with low amplitudes of
pulsations, and pulsating with nonradial modes (e.g. Fox Machado et
al. 2006).  Nonetheless, more precise information about its
evolutionary stage is needed for a more conclusive study.

A summary of the observations which led to the discovery and
characterization of the new $\delta$ Scuti star HD 207331 has been
presented. The star shows complicated pulsations as do most of the
$\delta$ Scuti stars.  To date our observations represent the most
extensive work on HD 207331.

\subsection{Future Work}
For the future, we will make spectroscopic observations of HD 207331
to obtain an accurate MKK spectral classification to fix more
exactly the basic physical parameters of this star in order to be
able to make a reliable seismic modeling.  Also, spectroscopic
observations will be carried out to test whether this $\delta$ Scuti
star rotates as rapidly as we might expect.  In addition, more
differential photometric observations, better distributed in time,
are needed to better understand this interesting object.

\acknowledgements This work has received financial support \\from
the UNAM via grant IN114309. WJS acknowledges financial support from
CONACyT by way of grant 49434-F. Special thanks are given to the
technical staff and night assistants of the San Pedro M\'artir and
Teide Observatories.  This research has made use of the SIMBAD
database operated at the CDS, Strasbourg (France). The IAC-80
telescope is operated by the Instituto de Astrof\'{\i}sica de
Canarias in the Observatorio del Teide.


\begin{thebibliography}{}

\bibitem{1993} Breger, M., et al. 1993, A\&A, 271, 482
\bibitem{2002} Fox Machado, L., et al. 2002, A\&A, 382, 556
\bibitem{2007} Fox Machado, L., et al. 2007, AJ, 134, 860
\bibitem{2008} Fox Machado, L., et al. 2008, CoAst, 156, 27
\bibitem{2006} Fox Machado, L., et al. 2006, A\&A, 446, 611
\bibitem{2005} Lenz, P., \& Breger, M. 2005, CoAst, 146, 53
\bibitem{2002} Li, Z.~P., et al. 2002, A\&A, 395, 873
\bibitem{1997} Perryman, M.~A.~C., et al. 1997, A\&A 323, L49
\bibitem{2008} Schuster, W.~J.,  Ochoa, J., Zurita, C., Fox Machado, L., 2008, IBVS, 5900,
8





\end{thebibliography}
\end{document}